\documentclass[aps,twocolumn,superscriptaddress]{revtex4-1}

\usepackage{graphicx}

\usepackage{amsmath}
\usepackage{amsfonts}
\usepackage{amssymb}
\usepackage{amsthm}
\usepackage{color}

\usepackage{dsfont}

\usepackage{hyperref}

\newcommand{\ket}[1]{|{#1}\rangle}
\newcommand{\bra}[1]{\langle{#1}|}

\newcommand{\beq}{\begin{equation}}
\newcommand{\eeq}{\end{equation}}

\newcommand{\COH}{\textup{coh}}
\newcommand{\SB}{\textup{SB}}

\newcommand{\EFF}{\textup{eff}}

\begin{document}

\author{C. Arenz} 
\affiliation{Frick Laboratory, Princeton University, Princeton, NJ 08544, USA}

\author{A. Metelmann} 
\affiliation{Department of Electrical Engineering, Princeton University, Princeton, NJ 08544, USA}
\affiliation{Dahlem Center for Complex Quantum Systems and Fachbereich Physik, Freie Universit\"{a}t Berlin, 14195 Berlin, Germany}

\title{Emerging unitary evolutions in dissipatively coupled systems}

\date{\today}

\begin{abstract}
Having a broad range of methods available for implementing unitary operations is crucial for quantum information tasks. We study a dissipative process commonly used to describe dissipatively coupled systems and show that the process can lead to pure unitary dynamics on one part of a bipartite system, provided that the process is strong enough. If suitably engineered, the process allows to implement generic unitary operations. In fact, we show within the framework of quantum control theory that the dissipative process can turn the system of interest into a system capable of universal quantum information task. We characterize the time scales necessary to implement gates with high fidelity through the dissipative evolution. The considered dissipative evolution is of particular importance since it can be engineered in the laboratory in the realm of superconducting circuits. 
Based on a reservoir that is formed by a lossy microwave mode we present a detailed study of how our theoretical findings can be realized in an experimental setting.    
\end{abstract}

\maketitle

\section{Introduction}
%
The implementation of unitary operations lies at the heart of quantum information processing. Quantum simulators, quantum metrology as well as quantum computing schemes and in general state preparation rely on the ability to implement unitary gates with high accuracy. It is therefore highly desirable to have a broad range of methods at hand for meeting that task. 
Typically unitary gates are implemented through external pulses, such as tailored optical fields and microwave fields \cite{NielseChuang}. Methods from quantum control theory \cite{BookDalessandro} can be used to determine the set of operations that can be implemented, whereas optimal control theory provides the tools to calculate the corresponding pulses to implement such gates with high accuracy \cite{ControlRev1}; even allowing to implement gates in the shortest possible time \cite{OptimalControlSpeedLimit1}. Over the last years quantum reservoir engineering schemes \cite{CiracZoller, BlattZoller}, particularly dissipative state preparation \cite{KrausZoller} and dissipative quantum computing \cite{WolfCirac} turned out to be a valuable alternative to unitary gate designs. For instance, instead of implementing a sequence of gates in order to carry out some computation, the computational step is entirely encoded in a suitably engineered dissipative process. Moreover, a strong dissipative process can lead to pure unitary dynamics over a subspace that is robust against the process being considered \cite{Zanardi1, Zanardi2, Fraas}. Such decoherence free subspaces \cite{DCBook} can be used to implement gates in a noiseless manner \cite{Beige}, and, furthermore, combined with methods from control theory, can turn parts of a system into a system capable of universal quantum computational tasks \cite{Me}. However, identifying decoherence free subspaces and engineering dissipative processes yielding a unitary evolution on parts of the system remains challenging.

\begin{figure}[!h] \includegraphics[width=0.9\columnwidth]{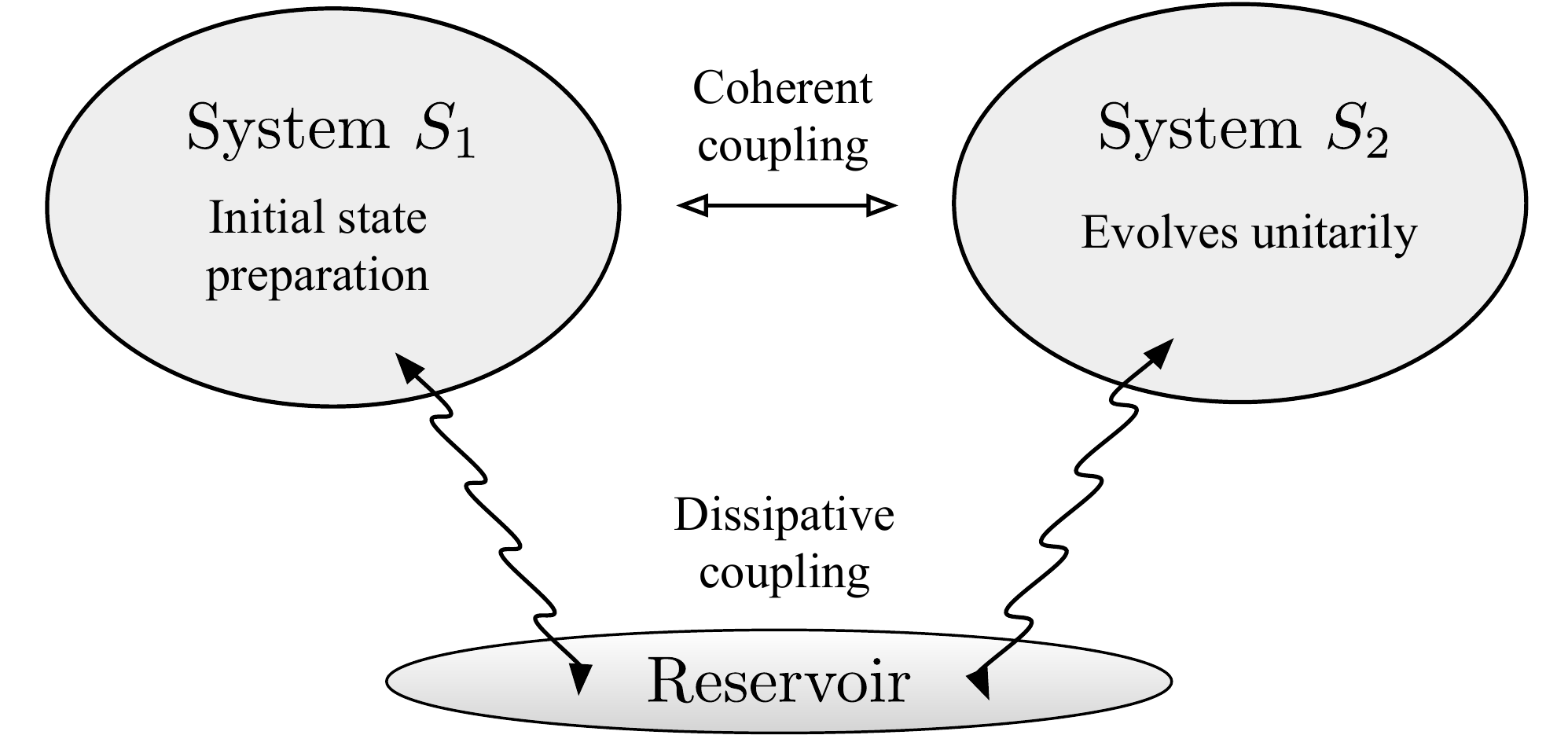}
  \centering
  \caption{\label{fig: schematicalRep} Schematic representation of two systems $S_{1}$ and $S_{2}$ that interact coherently described by the Hamiltonian \eqref{eq:coherenctinteraction} and in a dissipative way described by the Lindblad operator \eqref{eq:Lindbladorg} resulting from an interaction with a reservoir. For suitably engineered reservoirs the effect of the coherent interaction on $S_{2}$ can be enhanced or suppressed. In fact, if suitably engineered the dissipative process yields the same dynamics for $S_{2}$ as the coherent interaction would do (see Eq. \eqref{eq:equiv}). As such, for certain initial states of $S_{1}$, system $S_{2}$ evolves unitarily. If additionally system $S_{2}$ can be steered by some time dependent coherent process, the dissipative process can turn system $S_{2}$ into a system capable of universal quantum information tasks.}
\end{figure}

In this work we show that a dissipative process commonly used to describe dissipatively coupled systems $S_{1}$ and $S_{2}$  \cite{Krauter, Yingdang, Anja0, Schwartz, Liu}, can yield a purely unitary evolution on one system, say $S_{2}$.  In fact, we show that a dissipative process $\mathcal D$ described by the Lindblad operator  $L=\sqrt{\gamma} (A_{1}-i  \frac{\eta}{\gamma}B_{2})$ where $A_{1}$ and $B_{2}$ are hermitian operators on system $S_{1}$ and $S_{2}$, yields for large $\gamma$ the same dynamics on system $S_{2}$ as a coherent interaction between $S_{1}$ and $S_{2}$ would do. That is,  
\begin{align}
\label{eq:equiv}
\lim_{\gamma \to\infty}\text{tr}_{1}\{e^{\mathcal D t}(\rho_{1}\otimes\rho_{2})\}=\text{tr}_{1}\{U(t) (\rho_{1}\otimes \rho_{2})U^{\dagger}(t)\},
\end{align}
where $U(t)=\exp(-it H)$ is the overall unitary evolution generated by the Hamiltonian $H=\eta A_{1}B_{2}$, $\text{tr}_{1}\{\cdot\}$ denotes the partial trace over system $S_{1}$, and $\rho_{1}$ and $\rho_{2}$ are the initial states of both systems.  Consequently, when system $S_{1}$ is prepared in an eigenstate of $A_{1}$, the overall dissipative evolution yields for $\gamma\to\infty$ a purely unitary dynamics on system $S_{2}$ determined by $B_{2}$. 
Such being the case, as represented in Fig. \ref{fig: schematicalRep}, for two systems $S_{1}$ and $S_{2}$ undergoing an overall purely non-unitary evolution described by $\mathcal D$, system $S_{2}$ can evolve entirely unitarily provided the dissipative process is strong enough.  For the control properties of $S_{2}$ this implies that if system $S_{2}$ can be additionally steered by some time dependent fields, system $S_{2}$ can become fully controllable, and thus universal for quantum information tasks. We additionally provide criteria that characterize the time scales for implementing unitary gates with high fidelity through the dissipative process. 

The presented dissipative evolution is of particular importance since it can be realized in the laboratory using superconducting circuit architectures. Based on a reservoir that is formed by a strongly damped cavity mode, similar to the setting in \cite{Zoller, Rabl, Anja1, Anja2}, we discuss in detail how our theoretical findings can be experimentally realized.

%
\section{Dissipatively and coherently coupled systems}
%
In order to study the interplay between coherently and dissipatively coupled systems we considering two finite dimensional systems $S_{1}$ and $S_{2}$ that interact coherently described by the Hamiltonian 
\begin{align}
\label{eq:coherenctinteraction}
H_{\text{coh}}  = g A_{1}B_{2},	
\end{align}
with $A_{1}=A\otimes \mathds{1}_{S_{2}}$ and $B_{2}=\mathds{1}_{S_{1}}\otimes B$ being hermitian operators acting only non-trivially on system $S_{1}$ and $S_{2}$, respectively,  and $g$ being the interaction strength. The two systems are additionally coupled through a reservoir described by the Lindblad operator 
\begin{align}
\label{eq:Lindbladorg}
 L = \sqrt{\gamma}\left( A_{1} - \frac{\eta}{\gamma}  e^{i\phi} B_{2}\right),
 \end{align}
so that the evolution of the total system is given by the Lindblad master equation 
 \begin{align}
 	\dot{\rho}(t)=-i[H_{\text{coh}},\rho(t)]+\mathcal D[L](\rho(t)),
 \end{align}
 where $\rho$ is the state of the total system and $\mathcal D[L](\rho)=L\rho L^{\dagger}-\frac{1}{2}(L^{\dagger}L\rho+\rho L^{\dagger}L)$ with $L$ given by Eq.~\eqref{eq:Lindbladorg} is the Lindbladian describing the dissipative process. Here $\gamma$ denotes the rate associated with the dissipative process for system $S_{1}$,
while for system $S_{2}$ the corresponding rate is $\eta^2/\gamma$  $(\eta < \gamma)$, reflecting the asymmetric coupling to the dissipation. At this stage the Lindblad operator is introduced with a general phase $\phi$.
 The master equation then takes the form  
 \begin{align}
 \label{eq:masterform}
  \dot{\rho}(t)=& -i [ H_{\text{coh}},\rho(t)] + \gamma  \mathcal D[A_{1}](\rho(t)) 
                  +   \frac{\eta^2}{\gamma}  \mathcal D[B_{2}](\rho(t))
                 \nonumber \\ &
                  -  \mathcal K(\rho(t)) + \eta \cos(\phi) \{A_{1}B_{2},\rho(t) \},	
 \end{align}
with $\{\cdot,\cdot\}$ being the anti commutator and $\mathcal K(\rho(t))=\eta e^{i\phi}B_{2}\rho(t)A_{1}+\text{h.c.}$.	
From Eq.~\eqref{eq:masterform} we immediately see that for $\gamma\to\infty$, to which we refer to as the \emph{strong damping limit}, the dissipative term $\mathcal D[B_{2}]$ on system $S_{2}$ vanishes. Moreover, we note that the process $\mathcal D[A_{1}]$ does not effect system $S_{2}$. As we will see below, in the strong damping limit the term $\mathcal K$ can enhance or suppress the coherent interaction depending on the phase $\phi$ \cite{Anja1, Anja2}, as well as can give rise to pure unitary dynamics for system $S_{2}$.
     
The evolution of the state $\rho_{2}(t)$ of system $S_{2}$ is given by tracing over system $S_{1}$, i.e., $\dot{\rho}_{2}(t)=\text{tr}_{1}\{\dot{\rho}(t)\}$. If we evaluate the partial trace in the eigenbasis $\{\ket{\phi_{j}^{(a)}}\}$ of $A$ with corresponding eigenvalues $\lambda_{j}^{(a)}$ we find 
\begin{align}
\label{eq:dynamicsSys2}
\dot{\rho}_{2}(t) & =       - i \sum_{j}  \lambda_{j}^{(a)} [(g + \eta \sin(\phi)) B, \bra{\phi_{j}^{(a)}}\rho(t)\ket{\phi_{j}^{(a)}}]
            \nonumber \\ &
                            +  \frac{\eta^2}{\gamma}   \mathcal D[ B ](\rho_{2} (t)) ,
\end{align}
from which we see that, depending on $\phi$ and $\eta$, the dissipative interaction can enhance or suppress the effect of the coherent interaction on system $S_{2}$. Analogously, the evolution of the state $\rho_{1}(t)$ of system $S_{1}$ is governed by 
\begin{align}
\label{eq:dynamicsSys1}
\dot{\rho}_{1}(t) & =       - i \sum_{j}  \lambda_{j}^{(b)}      [(g  - \eta \sin(\phi)) A, \bra{\phi_{j}^{(b)}}\rho(t)\ket{\phi_{j}^{(b)}}]
            \nonumber \\ &
                            + \mathcal \gamma D[A](\rho_{1}(t)),	
\end{align}
where $\{\ket{\phi_{j}^{(b)}}\}$ is the eigenbasis of $B$ with corresponding eigenvalues $\lambda_{j}^{(b)}$. 
Notice the sign difference in the commutator part of Eq. \eqref{eq:dynamicsSys2} and Eq. \eqref{eq:dynamicsSys1}, which leads to uni-directional coherent dynamics due to the dissipative process $\mathcal D$.
This matches nicely the recipe introduced in Refs.~\cite{Anja1, Anja2}, where the balancing of a coherent and dissipative process can break the symmetry of reciprocity, rendering an interaction between two systems in a uni-directional fashion. For instance, for $\phi=\pi/2$ and $\eta = g$ the commutator part present in the dynamics of system $S_{1}$ vanishes, whereas for system $S_{2}$ the part coming from the coherent interaction \eqref{eq:coherenctinteraction} between both systems is enhanced due to the dissipative process.
In addition, under the directionality conditions $\phi=\pi/2$ and $\eta = g$ the full master equation resembles the one obtained from cascaded quantum systems theory \cite{BookGardinerZoller,Carmichael1},
i.e., the remaining (uni-directional) coupling yields 
\begin{align} 
  \dot{\rho}(t) \sim&    
                   \; i \eta \left\{  \left[  A_{1} \rho(t), B_{2}\right]     +   \left[ \rho(t) A_{1} , B_{2} \right]  \right\}    ,	
\end{align}
affecting only system $S_{2}$. However, an important difference to cascaded quantum systems theory here is, that one does not require a chiral information transfer via a waveguide to realize such a uni-directional interaction.

We proceed by focusing on the dynamics of system $S_{2}$.  One can easily check that in the strong damping limit the solution $\rho_{2}(t)=\mathcal E_{t}(\rho_{2}(0))$ to \eqref{eq:dynamicsSys2} is given by the (bistochastic) completely positive trace preserving (CPTP) map  
\begin{align}
\label{eq:unitaryCPTP}
\mathcal E_{t}(\cdot)=\sum_{j}p_{j}U_{j}(t)(\cdot)U_{j}^{\dagger}(t),	
\end{align}
where $p_{j}=\bra{\phi_{j}^{(a)}}\rho_{1}(0)\ket{\phi_{j}^{(a)}}$ with $\rho_{1}(0)$ being the initial state of system $S_{1}$ and the unitaries are given by 
$U_{j}(t)=\exp(-it\lambda_{j}^{(a)}(g+\eta\sin(\phi))B)$. For $\phi=\pi/2$ we have $U_{j}=\exp(-i(\eta +g)\lambda_{j}^{(a)}B)$ so that the effect of the coherent interaction is enhanced, thereby establishing for $\eta=g$ the equivalence expressed in \eqref{eq:equiv}. In the case where $A=\mathds{1}_{S_{1}}$ we can already see from the form of the Lindblad operator \eqref{eq:Lindbladorg} that $\mathcal E_{t}(\cdot)=U(t)(\cdot)U^{\dagger}(t)$ with $U(t)=\exp(-it(g+\eta\sin(\phi))B)$. For generic hermitian operators $A$ the preparation of system $S_{1}$ in an eigenstate of $A$ yields, up to a modification of $B$ in $U$ by the eigenvalue $\lambda_{a}$, the same unitary map. Furthermore, if in addition to $\mathcal D$ system $S_{2}$ is subject to some (possibly time dependent) coherent process 
$H_{2}(t)=\mathds{1}_{S_{1}}\otimes H(t)$, in the strong damping limit the dynamics of the state $\rho_{2}(t)$ of system  $S_{2}$ is governed by the von Neumann equation \begin{align}
\label{eq:vonNeumannSysB}
\dot{\rho}_{2}(t) = - i [\lambda_{a} (g + \eta\sin(\phi))B + H(t), \rho_{2}(t)].	
\end{align} 

In summary, for a suitable choice of the phase in the dissipative process $\mathcal D$, the strong damping limit enhances or suppresses the effect of the coherent interaction on one part of the bi partite system. Both, the strong damping limit of the dissipative process given by \eqref{eq:Lindbladorg} and the coherent process \eqref{eq:coherenctinteraction} independently yield the same CPTP map \eqref{eq:unitaryCPTP} for system $S_{2}$. We remark here that the same result can be obtained using a perturbative treatment \cite{Victor}. If we treat the $B_{1}$ term in $L$ as a perturbation to $A_{1}$, the unperturbed $\mathcal D[A_{1}]$ effectively yields a (projected) evolution over the decoherence free subspaces of $\mathcal D[A_{1}]$. As shown in \cite{Victor}, if we prepare the system in decoherence free subspace, the evolution over this subspace can be purely unitary.     

 In order to investigate the dynamics in more detail, we henceforth focus on the dissipative dynamics given by $\mathcal D[L] $ only, where we chose $\phi=\pi/2$ such that  
\begin{align}
\label{eq:lindbladop}
L=\sqrt{\gamma}\left(A_{1}-i \frac{\eta}{\gamma} B_{2}\right).	
\end{align}
We proceed with discussing a few implications of the previous observations. First of all we trivially see that a generic coherent evolution can be created on system $S_{2}$ through the dissipative process $\mathcal D$ . For instance, in the case where system $S_{2}$ is given by two non interacting spins, choosing $B=\sigma_{z}\otimes \sigma_{z}$ induces in the strong damping limit a coherent Ising type interaction.  Clearly, the challenge remains to engineer dissipative processes of the form $\mathcal D$ containing two body or many body interaction terms. Before we address this potential issue by providing a concrete experimental realization based on an engineered reservoir, we want to discuss in the context of quantum control theory how $\mathcal D$ can turn the system $S_{2}$ into a system capable of universal quantum information tasks.

\subsection{Universal control}
%
In general the aim of quantum control theory is to steer a quantum system towards a desired target by using a set of  suitably tailored classical \emph{control} fields $\{f_{k}(t)\}$. The total Hamiltonian describing the system reads $H(t)=H_{0}+ H_{c}(t)$ where the control typically enters in a bilinear way through $H_{c}=\sum_{k=1}^{n}f_{k}(t)H_{k}$. We refer to $H_{0}$ as the \emph{drift Hamiltonian} and to $\{H_{1},\cdots, H_{n}\}$ as the set of \emph{control Hamiltonians}. The system is said to be \emph{fully controllable} if every unitary transformation $U_{g}\in\text{SU}(d)$ (for traceless Hamiltonians) with $\text{SU}(d)$ being the group of unitary $d\times d$ matrices with determinant one can be implemented through shaping the control fields $f_{k}(t)$. It is known that every unitary operation in the closure of the dynamical Lie group $e^{\mathfrak{L}}$ can be implemented with arbitrarily high precision, with $\mathfrak{L}=\text{Lie}(i H_{0},i H_{1},\cdots, i  H_{n})$ being the real Lie algebra formed by real linear combinations of the drift and the control Hamiltonians and of their iterated commutators \cite{BookDalessandro}. If  $\mathfrak{L}=\mathfrak{su}(d)$, where $\mathfrak{su}(d)$ is the special unitary algebra, the system is said to be fully controllable. That is, every unitary can be implemented up to a global phase arbitrarily well. We remark here that operator controllability implies pure state controllability, i.e., every pure state can be prepared given that the system was initially prepared in a pure state. The dimension of the dynamical Lie algebra $\text{dim}(\mathfrak{L})$ characterizes how \emph{complex} the driven evolution can be \cite{DanielZeno}, and for a fully controllable system of dimension $d$ we have $\text{dim}(\mathfrak{L})=d^{2}-1$. In \cite{Me} it has been shown that a strong dissipative process exhibiting a decoherence free subspace can substantially change the dimension of the dynamical Lie algebra, even turning the system into a fully controllable ones. However, this effect critically relies on the ability to arbitrarily control two body interactions, and, moreover, the increase in $\text{dim}(\mathfrak{L})$ is limited by the dimension of the decoherence free subspace being considered.

In contrast, the strong dissipative process that is determined by the Lindblad operator \eqref{eq:lindbladop} offers a generic procedure for turning a quantum system through dissipation into a fully controllable ones and increasing the dimension of the dynamical Lie algebra arbitrarily. Suppose system $S_{2}$ is in addition to the overall acting $\mathcal D$ subject to some time varying controls, i.e., the time dependent Hamiltonian in \eqref{eq:vonNeumannSysB} is given by $ H(t)=\sum_{k=1}^{n}f_{k}(t) H_{k}$. Then, in the strong damping limit ($\gamma\to \infty$) the unitary operations that can be implemented on system $S_{2}$ are determined by the dynamical Lie algebra 
\begin{align}
\mathfrak{L}_{S_{2}}=\text{Lie}(iB,i H_{1},\cdots,i H_{n})	.
\end{align}
In the strong damping limit the hermitian operator $B$ given through the Lindblad operator $L$ takes the role of the drift Hamiltonian $H_{0}$. Thus the dynamical Lie algebra for system $S_{2}$ can be substantially different in the presence of the strong dissipative process $\mathcal D$. For instance, in the case of a single control Hamiltonian $H_{1}$ on system $S_{2}$ the dynamical Lie algebra is just one dimensional if dissipation is absent. Now, if $L$ can be engineered in such a way that $B$ generates together with $H_{1}$ the full algebra, i.e., $\mathfrak L_{S_{2}}=\mathfrak{su}(d)$ such that the dynamical Lie algebra has increased from 1 to $\text{dim}(\mathfrak L_{S_{2}})=d^{2}-1$, system $S_{2}$ is turned into a fully controllable system only due to the dissipative process. 
 There are several examples of pairs of Hamiltonians generating the full algebra, for example \cite{Carenzspinstar, FullControlSingleAcc}, and, moreover it can be shown that almost all pairs (but a set of measure zero) do the job \cite{Altafini}. Thus, system $S_{2}$ becomes for almost all choices of $B$ and $H_{1}$ fully controllable.

\subsection{Timescales}
%
So far we have studied the time evolution of system $S_{2}$ in the strong damping limit, i.e. for $\gamma \to\infty$. Now we want to investigate the effect of a finite $\gamma$ on the fidelity for preparing a state. We consider the fidelity error $\epsilon=1-F$ where $F=\bra{\psi_{G}}\rho_{2}(t)\ket{\psi_{G}}$ is the fidelity for preparing a pure state $\ket{\psi_{G}}$ and $\rho_{2}(t)$ is the state at time $t$ of system $S_{2}$. We assume here that system $S_{2}$ was initially prepared in a pure state $\ket{\phi(0)}$ such that $\ket{\psi_{G}}=U\ket{\psi(0)}$ is prepared at time $t$ on system $S_{2}$ in the strong damping limit. 

We begin with the case for which no additional coherent term on system $S_{2}$ is present so that the time evolution of system $S_{2}$ is entirely determined by the Lindblad operator \eqref{eq:lindbladop}. Since all processes contained in $\mathcal D$ mutually commute with each other and assuming that the process $\mathcal K$ with $\phi=\pi/2$ prepares the state $\ket{\psi_{G}}$ at time $t$, the time evolution of system $S_{2}$ is given by $\rho_{2}(t)=\exp(\frac{\eta^{2}}{\gamma}\mathcal D[B_{2}]t)(\ket{\psi_{G}}\bra{\psi_{G}})$. Expanding $\ket{\psi_{G}}=\sum_{k}c_{k}\ket{n}$ in the eigenbasis $\{\ket{\phi_{n}^{(b)}}\}$ of $B$ with corresponding eigenvalues $\{\lambda_{n}^{(b)}\}$ the fidelity error then reads $\epsilon=1-\sum_{n, m}|c_{n}|^{2}|c_{m}|^{2}\exp(-\frac{t\eta^{2}}{2\gamma}(\lambda_{n}^{(b)}-\lambda_{m}^{(b)})^{2})$.	
We can conclude that we need
\begin{align}
\frac{\gamma}{\eta^{2}} \gg \frac{t}{2}\max_{n\neq m}(\lambda_{n}^{(b)}-\lambda_{m}^{(b)})^{2}, 
\end{align}	
in order to prepare the state $\ket{\psi_{G}}$ at time $t$ with high fidelity through the dissipative process. 

We proceed with the case in which system $S_{2}$ is additionally subject to some possibly time dependent coherent process described by $H_{2}(t)=\mathds{1}_{S_{1}}\otimes  H(t) $ where $ H(t)$ could for instance be of the form $H(t)=\sum_{k=1}^{n}f_{k}(t)H_{k}$. We saw in the previous  paragraph that in this case in the strong damping limit every unitary operation $U_{g}=e^{\Theta}$ with $\Theta\in\mathfrak{L}_{S_{2}}$ can be implemented on system $S_{2}$. Here we now want to study the fidelity error for finite $\gamma$ for preparing the corresponding state $\ket{\psi_{G}}=U_{g}\ket{\psi(0)}$ at time $t$. Because the relevant process do not necessarily commute anymore, and moreover the total generator is now time dependent, an exact expression for $\epsilon$ as before is not trackable anymore.  However, with details found in Appendix \ref{ref:fiderrorupbound} 
we can upper bound the fidelity error by 
\begin{align}
\label{eq:upperboundfiderror}
\epsilon\leq \frac{t\eta^{2}}{2\gamma}(\Vert B\Vert_{\infty}^{2}+\Vert B^{2}\Vert_{\infty}),	
\end{align}
where $\Vert \cdot \Vert_{\infty}$ is the standard operator norm.
 
Having discussed the theoretical properties of the dissipative process $\mathcal D$, we now turn to presenting an experimental realization of $\mathcal D$.

\section{Experimental realization}
%
In general, dissipation is trivially modeled by coupling the system of interest to a Markovian bath.
Information is then simply lost into this bath forever and the bath does not mediate any inner correlations in the system.
In contrast, engineered dissipation is a controlled form of dissipation, here the system of interest is 
coupled to a damped auxiliary system which mediates a manipulable dissipative process.
Engineering a non-local dissipative process of the form $ \mathcal D \left[ L \right] (\rho)$ between system $S_1$ and $S_2$,
with the jumpoperator $L$ being a combination of hermitian operators of $S_1$ and $S_2$, e.g. Eq.~(\ref{eq:Lindbladorg}), 
requires both systems to be coupled to a strongly damped auxiliary system in a coherent and controllable manner.   
The easiest form of such an auxiliary system is a damped mode $a$. Then the required coherent system-bath dynamics are described by a Hamiltonian of the form 
\begin{align}
H_{\textup{SB}} =  \;  \lambda_1  X_{\varphi_{1}}   A_{1}   + \lambda_2  X_{\varphi_{2}}   B_{2}   ,   
\end{align}
with $X_{\varphi_{n}} =    \left[   a    e^{-i \varphi_{n} } +   a^{\dag}   e^{i \varphi_{n} } \right] $ being   the quadrature  operators of the $a$-mode.
These quadratures $X_{\varphi_{n}}$ do not have to be orthogonal, but crucial is their relative phase $\varphi_{1} - \varphi_{2} $, 
which determines the phase in the resulting non-local jumpoperator $L$.
To have  $H_{\textup{SB}}$ generate the dissipative process $ \mathcal D \left[ L \right] (\rho)$, we couple mode $a$ to a Markovian bath with rate $\gamma_a$.   
In the case of strong damping, i.e., for $\gamma_a \rightarrow \infty$, the auxiliary mode can be adiabatically 
eliminated  \cite{Carmichael2,Anja3} and one is left with the dissipative process described by the Lindblad operator
\begin{align}
 L = \frac{2\lambda_1 }{\sqrt{\gamma_a}}  \;  \left[  A_{1} + \frac{\lambda_{2}}{\lambda_1}  e^{-i (\varphi_{1}- \varphi_{2}) }   B_{2}  \right]. 
\end{align}
Thus, an asymmetry in the coherent couplings $\lambda_{1 }$ and $\lambda_2$ translates directly to an asymmetry in the dissipative process between system $S_1$ and $S_2$.
And, as mentioned above, the relative phase of the quadratures allows for a finite phase in the non-local dissipator. 
Note that here $ H_{\textup{SB}}$ has to be a resonant interaction, i.e., if the auxiliary mode is detuned,
one generates an effective coherent coupling $  H_{\EFF} \sim A_{1} B_{2} $ between system $S_1$ and $S_2$ too.

In the following section we present a concrete example of how to realize a non-local dissipative process between a three-spin system.

\subsection{A superconducting circuit implementation}
%
  \begin{figure}[!h] \includegraphics[width=1.0\columnwidth]{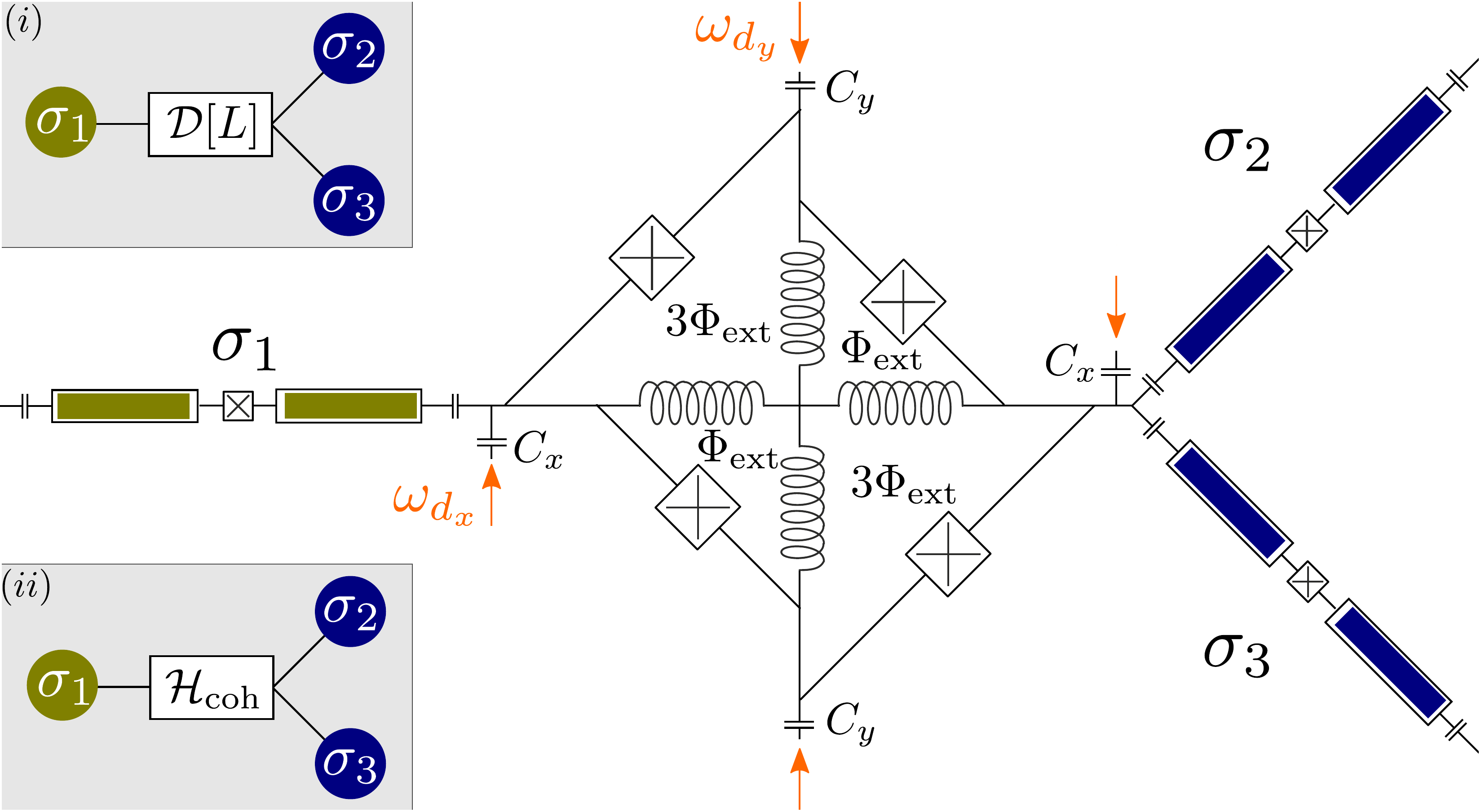}
  \centering
        \caption{\label{fig:ExpSketch}
                Three-qubit coupling via a Josephson ring modulator. 
	        The resulting interaction between the qubits can be  either of the dissipative form (i) $\mathcal D [L]$ with $L =  \sigma_1 +   \sigma_2   \sigma_3 $
	        or of the coherent form (ii) $  H_{\COH} =   \sigma_1   \sigma_2   \sigma_3 $. The nature of the interaction is determined by the external drives 
	        (see text for details).
	        }
\end{figure}

Over the last decade, the realm of superconducting circuitry \cite{Blais2004, Devoret2013,Gu2017} 
has experienced tremendous growth due to advances
in nanofabrication technologies, which in turn have led to an impressive progress in the development of quantum
technologies. Despite being macroscopic elements, i.e., on the length-scale of hundreds of nanometers, superconducting circuits 
behave quantum mechanically, as they can be designed to be well isolated from the environment. 
For a recent review please see Ref.~\cite{Gu2017}.

The basic toolbox of superconducting circuits utilized  for quantum simulation and quantum computation 
consists of linear and non-linear resonators, where the latter can be operated as artificial few-level atoms or qubits. 
Superconducting qubits are formed via the two lowest energy states of a nonlinear Kerr resonator. 
The nonlinearity of the resonator is crucial here for the design of the qubits, as it is accompanied with discrete energy levels which are not equally spaced
(in contrast to a linear oscillator). The nonlinear Kerr resonator can be realized by combining a linear LC-resonator circuit with 
a nonlinear and dissipation-less inductance: the Josephson junction. Once placed
into a low temperature environment these nonlinear Kerr resonators enter the quantum regime
and can be treated as artificial two-level systems aka qubits.
Mixing between multiple qubits can be accomplished via tunable couplers \cite{Chen2014, Lecocq2017, Roushan2017, Bergeal2010a}, 
and the read-out, manipulation, and control of the qubits can be realized via the coupling to the
discrete electromagnetic modes of quantum cavities or to the continuum of modes in a waveguide.

In this section we are going to discuss a concrete example on how to engineer a non-local dissipative process in a three-qubit system
based on a superconducting circuit architecture.
We like to stress that this is just one of many possible realizations and we choose the present
setup because it nicely illustrates that one circuit can provide the same type of coherent and dissipative non-linear process, with 
the difference that engineering the coherent interaction requires processes 
that are of higher order than the processes leading to  dissipative interactions.

The multi-qubit system we like to consider is formed by three nonlinear Kerr resonators, e.g., a transmission line intersected with a Josephson junction, 
which are operated in the low excitation and low dissipation regime. For a strong enough Kerr nonlinearity each resonator 
can be considered as an effective two level system, which we describe by the Pauli spin operators $ \sigma_{n}$, where $n=1,2,3$ labels each two level system. 
To realize a dissipative coupling between the qubits we aim for the situation that all three nonlinear resonators are coupled to the same dissipative reservoir, 
i.e., an auxiliary mode which is strongly damped via the coupling to a Markovian bath with rate $\gamma_{a}$. 
We focus on realizing a non-local jumpoperator $L$ of the form given in Eq.~(\ref{eq:Lindbladorg}) with $A_{1}=\sigma_1$ and $B_{2}=\sigma_{2} \sigma_{3}$.
As discussed above, such a nonlinear dissipative process is realized via the system-bath Hamiltonian
\begin{align}\label{Eq.SystemBathHam} 
 H_{\textup{SB}} =& \;  \frac{\sqrt{\gamma \gamma_a}}{2}  \;   \left[  X_{\varphi_{1}} \sigma_1 + \frac{\eta}{\gamma} X_{\varphi_{2}} \sigma_2  \sigma_{3} \right]  , 
\end{align}
with $\varphi_{2}- \varphi_{1} = \pi + \phi$, and $\phi,\eta,\gamma$ as introduced in Eq.~(\ref{eq:Lindbladorg}).
We leave the spin-component, i.e., $\sigma_{n} \rightarrow \sigma_{n}^{x,y,z}$, unspecified for now. 
To realize the interaction in $  H_{\textup{SB}}$ we use a Josephson ring modulator (JRM)\cite{Bergeal2010a}, which consists
of four identical Josephson junctions embedded in a ring geometry. This device provides three-wave mixing between its three spacial mode amplitudes $\phi_{x,y,z}$ and was originally 
developed for quantum-limited amplification of weak signals \cite{Bergeal2010b}. The whole circuit is sketched in Fig.~\ref{fig:ExpSketch} and can be modeled via the Hamiltonian
\begin{align}
  H  =&  
       \;   H_{0}  
      + \sum_{m = x,y,z } \sum_{n = 1}^{3}  g_{nm}     \left[  d_{m}  \sigma^{+}_n    +  d_m^{\dag} \sigma^{-}_n    \right]  
      + V_{\textup{JRM}},
\end{align}
where $  H_{0}$ contains the free energy of the two-level systems and the JRM modes $ \phi_{m} = \phi_{0,m} (   d_{m} +  d_{m}^{\dag} ) $ where $\phi_{0,m}$ denotes the standard deviation of the zero-point flux fluctuation for the JRM mode $  \phi_{m}$.
The second term describes excitation exchange between the qubits and the JRM modes with
interaction strength $g_{nm}$, which  depend  on the design of the coupling capacitors $C_{m}$, cf. Fig.~\ref{fig:ExpSketch}.
$V_{\textup{JRM}}$ denotes the mixing potential 

\begin{align}
 V_{\textup{JRM}} =& - E_{J} \sum_{\pm}
                           \left[  \cos \frac{  \phi_{x} \mp   \phi_{y}}{2\phi_0}  \cos \frac{ 2 (2 \pm 1) \phi_{\textup{ext}} \pm  \phi_{z} }{ 2 \phi_0}    \right]     
\end{align}

for the spacial mode amplitudes $\phi_{x,y,z}$ realized via the JRM and  
the latter potential is tunable via the external flux $\phi_{\textup{ext}}$.  
$E_J$ denotes the Josephson energy, which is assumed to be identical for all four junctions,
and $\phi_{0} = \hbar/2e$ corresponds to the reduced flux quantum.
We choose a design where the JRM loop is shunted with linear inductors as depicted in Fig.~\ref{fig:ExpSketch}.
For simplicity we neglect the frequency shifts associated with the potential energies of the inductors.
The resulting inner loops of the JRM are asymmetrical biased, i.e., with an external flux $\phi_{\textup{ext}} (3 \phi_{\textup{ext}})$ for the small (big) loops.
Such kind of setup was proposed earlier to realize tunable multi-body interactions employed to protect quantum information in cat-code approaches \cite{Mirrahimi} 
and for quantum annealing protocols \cite{Puri}. 

For our purpose we set $\phi_{\textup{ext}} = \pi/4 \phi_0$
and assume that the x-mode and y-mode are externally driven by multiple pump tones. 
This external driving ensures that the otherwise far off  resonant nonlinear processes are enforced.
For now we do not further specify the involved driving frequencies, but make the classical approximation 
$ \phi_{x}  \phi_{y} \rightarrow 4 \phi_{0,x} \phi_{0,y} \alpha_{x}  \alpha_{y} \mathcal M (t) $, where $|\alpha_{n}|^2$ denotes 
the average photon number in the $n$-mode induced by the external drives, and
the time-dependent modulation is given by  
 \begin{align} \label{Eq.Modulation}
   \mathcal M (t)  =& \; \prod_{n =x,y} \sum_{m} \cos(\omega_{{n,m}}^{d} t + \phi_{n,m}),
 \end{align}
with $m$ drives on each mode  with frequencies $\omega_{n,m}^{d}$; crucially, these drives are associated with the phases $\phi_{n,m}$. 

Expanding the JRM mixing potential yields
\begin{align}\label{Eq.JRMpotential}
  V_{\textup{JRM}}   \approx & \;   \frac{E_J  \alpha_{x}^{\prime} \alpha_{y}^{\prime}}{ 2 \sqrt{2} } \;   \mathcal M(t) \;
                           \left[  \frac{ \phi_{z} }{ \phi_0} -   \frac{ \phi_{z}^2 }{4\phi_0^2} -  \frac{  \phi_{z}^3 }{24\phi_0^3}  \right],  
\end{align}
with $\alpha_{n}^{\prime} =   \alpha_{n} \phi_{0,n} /\phi_0 $.  
In what follows, the z-mode is going to be our auxiliary mode 
and the choice of the drive frequencies  will determine which interactions are resonant in the three-spin system. 
The frequencies of the circuit should be engineered such that all three qubits are dispersively coupled to
the z-mode. In this regime we can perform the Schrieffer-Wolff transformation 
\begin{align} 
& 
  H^{\prime}  = e^{-  S}    H e^{  S} ,  
   \hspace{0.1cm}
   S =        \sum_{n = 1}^{3}  \lambda_{nz}     \left[  d_z^{\dag}  \sigma^{-}_n   -   d_{z}  \sigma^{+}_n        \right]   ,
\end{align}
where $\lambda_{nz} =  g_{nz} /\Delta_{nz}$ and $\Delta_{nz}$ denotes the detuning of qubit $n$ with respect to the z-mode.
In the dispersive limit $\lambda_{nz}$  is small and we only keep terms up to  second order in $\lambda_{nz}$.
In addition, we apply a rotating wave approximation to eliminate fast rotating terms. The remaining effective interaction yields
 \begin{align} \label{Eq.SBHamXXXcoupling}
   H_{\EFF}  = &    
       -      \Lambda  \; \mathcal M(t) \left(   d_z +   d_z^{\dag} \right) 
        \left[        \lambda_{1z} \;    \sigma^{x}_1  
                 +     \beta \;    \sigma^{x}_2      \sigma^{x}_3 \right]  , 
 \end{align}
 with the coefficients  
 \begin{align}
    \Lambda =&    \frac{E_J  \alpha_{x}^{\prime} \alpha_{y}^{\prime}}{ \sqrt{2} } \; \frac{  \phi_{z,0}^2}{4 \phi_0^2}   ,
              \hspace{0.2cm}
      \beta =   \lambda_{2z}  \lambda_{3z}  \frac{\phi_{z,0}}{2\phi_0}   .
\end{align}
The effective interaction Hamiltonian $ H_{\EFF}$ is close to the desired form, cf. Eq.~\ref{Eq.SystemBathHam}, but it is still time-dependent
through the modulation $\mathcal M(t)$.  

This time-dependence can be omitted by moving into the right rotating frame and choosing the appropriate driving frequencies.
First we move into in interaction frame with respect to the modified free Hamiltonian $H_{0}^{\prime}$, i.e.,
the free energy part of the Hamiltonian after the Schrieffer-Wolff transformation has been performed.
This unitary operation gives us for the spin-operators
$\sigma_{n}^{x} \rightarrow \sigma_{n}^{+} e^{+ i \Omega_{n} t }+ \sigma_{n}^{-} e^{- i \Omega_{n} t }$ 
and the z-mode operator  $d_{z} \rightarrow d_{z}  e^{- i \omega_{z} t }$,
where $\Omega_{n} (\omega_z)$ is the (shifted) frequency of qubit $n$ (z-mode). 
Inserting these expressions into the interaction Eq.~\ref{Eq.SBHamXXXcoupling}, 
we can identify the required driving frequencies, e.g., the processes $ d_{z} \sigma_{n}^{\pm}$
oscillate in this frame with $(\omega_{z} \pm \Omega_{n})$, thus choosing the external modulation
at these frequencies renders these processes resonant. Overall, we find six 
modulation frequencies to obtain the desired operators $A_{1}$ and $B_{2}$:
  
\begin{align}\label{Eq.PumpToneFrequencies}
 \omega_{1,\pm} =   \; \omega_{z} \pm \Omega_{1}
 \hspace{2.0cm} &\Rightarrow \hspace{0.5cm} 
  A_{1} = \sigma_{1}^{x},
  \nonumber \\ 
    \left.
      \begin{array}{ll} 
          \omega_{2,\pm} =   \; \omega_{z} \pm \left(\Omega_{2} + \Omega_{3} \right)\\
          \omega_{3,\pm} =   \; \omega_{z} \pm \left(\Omega_{2} - \Omega_{3} \right)
                \end{array}
  \right\}
                \hspace{0.5cm} &\Rightarrow \hspace{0.5cm} 
                B_{2} = \sigma_{2}^{x} \sigma_{3}^{x}
                .
\end{align}  
Luckily, the six frequencies $\omega_{n,\pm}$ are asymmetric and anti-symmetric combinations of four basic tones,
thus an appropriately chosen four-tone driving of the modes $x$ and $y$ is sufficient to produce these six tones, cf. Eq.~\ref{Eq.Modulation} with $\omega_{{n,m}}^{d}$ as the external drive frequencies.
To obtain the required modulation frequencies $\omega_{m,\pm},  (m =1,2,3)$ it is sufficient to 
drive the x-mode with one tone at $\omega_{x,1}^{d} = \omega_{z}$ 
and the y-mode at three different frequencies: $\omega_{y, m }^{d} = \Omega_1, \Omega_{2} \pm \Omega_{3}$.
Setting these frequencies into Eq.~\ref{Eq.Modulation} and applying basic trigonometric product rules results 
in the modulation  $\mathcal M  (t) =  \mathcal M_{+} (t) +  \mathcal M_{-} (t)$ with
\begin{align} 
 \mathcal M_{\pm} (t)  =& \;  \frac{1}{2} \; \sum_{m = 1 }^{3} \cos[ (\omega_{z} \pm \omega_{y, m}^{d} ) t + \phi_{m,\pm}] ,
\end{align}
and with the definition $\phi_{m, \pm} = \phi_{y, m} \pm \phi_{x,1}$.Combining this modulation with the interaction given in Eq.~\ref{Eq.SBHamXXXcoupling}, performing a rotating wave approximation, and
setting $\phi_{1,+} = \phi_{1,-} \equiv \phi_1  $ and $\phi_{2,\pm} =   \phi_{3,\pm}  \equiv \phi_{2}  $, leaves us with the resonant (time-independent) terms: 
\begin{align} \label{EqFinalHSB}
  H_{\EFF}   \approx   &   \;  
                  - \frac{\Lambda}{4} \;    \left( \lambda_{1z}  X_{\phi_{1}}   \hat \sigma^{x}_1  +   \beta \; X_{\phi_{2}}   \hat \sigma^{x}_2  \hat \sigma^{x}_3 \right) 
                  \equiv H_{\SB}^{\prime}, 
\end{align}
which is of the desired form for the system-bath interaction, cf. Eq.~\ref{Eq.SystemBathHam}. 
In a last step we assume that the z-mode is strongly damped with rate $\gamma_z$, so we can adiabatically eliminate it.
With the mapping $\Lambda \lambda_{1z} = 2\sqrt{\gamma \gamma_z} $, $\eta/\gamma = \beta/\lambda_{1z} $ and $\phi_{2}- \phi_{1} = \pi + \phi$,
we obtain $L$ of the form given in Eq.~(\ref{eq:Lindbladorg}). 
%
\subsection{Reciprocal and nonreciprocal coherent dynamics}
%
The circuit design presented in the last subsection provides the dissipative process  
for system $S_1$ and $S_2$ described by the jump-operator  

\begin{align} \label{EqFinalDiss}
  L   = &  \; \frac{\Lambda  \lambda_{1z} }{2\sqrt{\gamma_z}} \;  
        \left(   \sigma_1^{x}    +   \frac{\beta}{\lambda_{1z} }  e^{-i(\phi_1 - \phi_2)} \sigma_2^{x} \sigma_{3}^{x} \right) ,    
 \end{align} 
resulting in the effective decay rate $\gamma = \Lambda^2 \lambda_{1z}^2/(4\gamma_z)$,
which can be adjusted by varying the pump-amplitudes. 

As discussed in Sec.~II, in the \textit{strong damping limit} one can create effective coherent dynamics
for system $S_2$. Crucially, we have to make a distinction here in terms of what we consider strong damping.
One the one side we have the dissipation of the z-mode associated with the rate $\gamma_z$. 
The latter has to be large to obtain $L$ out of  $  H_{\SB}^{\prime}$ given in Eq.~(\ref{EqFinalHSB}). On the other side, to
realize coherent dynamics in $S_2$ the \textit{engineered} dissipation has to be strong, i.e., we have to realize $\eta^2 /\gamma  \rightarrow 0$, cf. Eq.~(\ref{eq:dynamicsSys2}).
For the discussed experimental example we can extract the condition for the \textit{strong damping limit} as

\begin{align}
  \frac{\eta^2}{\gamma}  \Rightarrow   \left[  \gamma \;    \frac{\lambda_{2z}^2  \lambda_{3z}^2}{\lambda_{1z}^2}  \; \frac{\phi_{z,0}^2}{4\phi_0^2}    \right]  \rightarrow 0,
\end{align}

thus, one achieves this limit for the qubits in system $S_2$ in the deep dispersive regime 
where $\lambda_{2z}^2  \lambda_{3z}^2 \ll \lambda_{1z}^2$,  
as well as small effective decay rate $\gamma$.
The latter scales inversely with the decay rate of the z-mode, i.e., $\gamma \sim 1/\gamma_z$, thus the condition is in good agreement 
with the requirement of a strongly damped z-mode.
However, we see here that the \textit{strong damping limit} for this experimental realization
is rather a strong dispersive limit, where the hierarchy $\lambda_{2z}^2  \lambda_{3z}^2 \ll \lambda_{1z}^2$ is the crucial ingredient.

The \textit{strong damping limit} results in coherent dynamics of system $S_2$ as would have been obtained from  $  H_{\COH} =  g A_1 B_2   $.
As mentioned in Sec.~II, having both processes, the dissipative and the coherent one, enables us to render the system directional.
The introduced circuit architecture allows as well the realization of a coherent interaction of the form given in Eq.~(\ref{eq:coherenctinteraction}). 
In the dispersive regime we obtain the process  
\begin{align}
     H_{\COH}^{\prime}  =&    
   -  \lambda \; \mathcal M(t)   \;   \hat \sigma^{x}_1  \hat \sigma^{x}_2  \hat \sigma^{x}_3  , 
   \hspace{0.3cm}
  \lambda = \Lambda  \lambda_{1z}   \beta ,
\end{align}
which originates from the cubic term in the $V_{\textup{JRM}}$ potential given in Eq.~(\ref{Eq.JRMpotential}). 
This coherent interaction is a third order process in the dispersive limit, i.e., it scales with $\lambda_{1z} \lambda_{2z} \lambda_{3z}$,
in contrast to this, for the (equivalent) dissipative process the second order of the dispersive limit was sufficient. 
Note, considering the cubic term in the potential would in principle require to perform the Schrieffer-Wolff transformation up to the third order as well, 
an additional step we have omitted here.

For now we just want to briefly illustrate how the introduced circuit architecture can realize $ H_{\COH}^{\prime}$.
The required drive frequencies are obtained by making the substitution $\omega_{z} \rightarrow \Omega_1$ in Eq.~(\ref{Eq.PumpToneFrequencies}).
Thus, we can still work with the same basic tones and just add another drive to the  x-mode at $\omega_{x,2}^{d} = \Omega_1$.
The total modulation becomes $\mathcal M_{\textup{tot}} (t) = \mathcal M (t) + \mathcal M_{+}^{\COH}  (t) +  \mathcal M_{-}^{\COH}  (t) $
with 
\begin{align} 
 \mathcal M_{\pm}^{\COH} (t)  =& \;   \frac{1}{2} \;  \sum_{m = 1 }^{3}  
                                  \cos[ (\Omega_{1} \pm \omega_{y,m}^{d} ) t + \theta_{m,\pm}]  ,
\end{align} 
and the phases $\theta_{m, \pm} = \phi_{y, m} \pm \phi_{y,1}$. 
Note, the modulation $\mathcal M_{\textup{tot}} (t)$ results as well in a unused tone at $2 \Omega_1$,
which should not drive any additional process if the involved resonances are designed appropriately.
For $\theta_{m, \pm} = \pi$ we obtain the coherent interaction $ H_{\COH}^{\prime} =  \lambda/4   \; \hat \sigma^{x}_1  \hat \sigma^{x}_2  \hat \sigma^{x}_3 .$

Combining now this coherent process $ H_{\COH}^{\prime}$, and the dissipative process 
$ \mathcal D \left[ L \right] (\rho)$ with the jumpoperator $L$ given in Eq.~(\ref{EqFinalDiss}) and $\phi_{2}- \phi_{1} = \pi + \phi$,
the interaction between system $S_{1}$ and $S_{2}$ becomes fully nonreciprocal under the conditions \cite{Anja1, Anja2}:
\begin{align}
 \phi = \pm \frac{\pi}{2},
 \hspace{0.5cm}
      \Lambda    =        \gamma_{z} ,
\end{align}
where the sign of the phase determines whether system $S_{1}$ or $S_{2}$ is affected by the dynamics of the respective other system.
For example for $\phi = \pi/2$ system $S_{2}$ performs enhanced coherent dynamics while system $S_{1}$ is not affected, 
cf. Eq.~(\ref{eq:dynamicsSys2}) and Eq.~(\ref{eq:dynamicsSys1}).
\\[2mm]

\section{Conclusion}
%
We have analyzed two systems $S_{1}$ and $S_{2}$ that interact in coherent and a dissipative way mediated though a reservoir. We showed that the dissipative process can enhance or suppress the effect of the coherent interaction on system $S_{2}$. In fact, if suitably engineered, the dissipative process has the same effect on $S_{2}$ as the coherent interaction. Consequently, for certain initial states of system $S_{1}$, system $S_{2}$ evolves unitarily. We have shown that if system $S_{2}$ can be additionally be steered by some time dependent fields, the dissipative process  can turn system $S_{2}$ into a fully controllable system which is capable of universal quantum information task, i.e., every unitary gate can be implemented on $S_{2}$. Furthermore, based on superconducting circuits, we have presented a scheme to engineer a reservoir that yields the desired dissipative process, as well as the "equivalent" coherent interaction. It is interesting to note that engineering coherent couplings requires processes that are of higher order than the processes leading to  dissipative couplings. Given the equivalence of both processes for the dynamics of system $S_{2}$, this suggest that engineering the desired dissipation may be more applicable to coherently control one part of a system.


{\em Acknowledgements.} -- 
The authors wish to thank Victor Albert for useful discussions. AM acknowledges funding by the  Deutsche Forschungsgemeinschaft through the Emmy Noether program (Grant No. ME 4863/1-1)
and the project CRC 910. 
%
%
\bibliographystyle{apsrev}
 

\onecolumngrid
\appendix

\section{Derivation of the upper bound for the fidelity error}
\label{ref:fiderrorupbound}
Here we derive the upper bound \eqref{eq:upperboundfiderror} for the fidelity error $\epsilon$ in the time dependent case described by the master equation
\begin{align}
\dot{\rho}(t)=\mathcal D[L](\rho(t))-i[H_{2}(t),\rho(t)],	
\end{align}
where the Lindblad operator $L$ is given by \eqref{eq:Lindbladorg} and $H_{2}(t)=\mathds{1}_{S_{1}}\otimes H(t)$ is a coherent time dependent process only acting non-trivially on system $S_{2}$, assume that no coherent interaction between $S_{1}$ and $S_{2}$ is present. We set the phase in \eqref{eq:Lindbladorg} to $\phi=\pi/2$ and we include the constant $\eta$ in the operator $B$ acting on $S_{2}$. We work in the frame rotating with $H(t)$. That is, we introduce the rotated state $\tilde{\rho}=V^{\dagger}(t)\rho V(t)$ with $V(t)=\mathcal T\exp(-i\int_{0}^{t}H(t^{\prime})dt^{\prime})$ so that the master equation in the rotated frame reads 
\begin{align}
\dot{\tilde{\rho}}(t)=\mathcal D[L(t)](\tilde{\rho}(t)),	
\end{align}
 where $L(t)=\sqrt{\gamma}(A_{1}-i\gamma^{-1}B_{2}(t))$ with $B_{2}(t)=V^{\dagger}(t)B_{2}V(t)$ so that
 \begin{align}
 \dot{\tilde{\rho}}(t)=\gamma \mathcal D[A_{1}](\tilde{\rho}(t))+\gamma^{-1}\mathcal D[B_{2}(t)](\tilde{\rho}(t))-i(B_{2}(t)\tilde{\rho}(t)A_{1}-A_{1}\tilde{\rho}(t)B_{2}(t)).
 \end{align}
If system $S_{1}$ is initially prepared in an eigenstate $\ket{a}$ of $A$ with corresponding eigenvalue $\lambda_{a}$, in the limit $\gamma \to\infty$ the dynamics of system $S_{2}$ is given by the unitary map $\tilde{\mathcal U}_{t}$ generated by $ \tilde{\mathcal H}_{t}(\cdot)=-i\lambda_{a}[B_{2}(t),\cdot]$. 
 We now want to study the effect of a finite $\gamma$ by upper bounding the fidelity error $\epsilon=1-F$ with $F=\bra{\psi_{G}}\rho_{2}(t)\ket{\psi_{G}}$ being the fidelity. We assume that system $S_{2}$ is initially prepared in a pure state $\ket{\phi(0)}$ such that in the limit $\gamma\to\infty$ the target state $\ket{\tilde{\psi}_{G}}$ (in the rotated frame) is prepared on system $S_{2}$. That is, if we assume that the initial state of the total system is given by 
 \begin{align}
 \rho(0)=	\ket{a}\bra{a}\otimes \ket{\psi(0)}\bra{\psi(0)},
 \end{align}
we have $\lim_{\gamma\to\infty}\text{tr}_{S_{1}}\left\{\mathcal Te^{\int_{0}^{t}dt^{\prime}\mathcal D[L(t^{\prime})]}(\rho(0))\right\}=\tilde{\mathcal U_{t}}(\rho(0))=\ket{a}\bra{a}\otimes \ket{\tilde{\psi}_{G}}\bra{\tilde{\psi}_{G}}$, whereas for finite $\gamma$ the state $\tilde{\rho}_{2}(t)$ in the rotated frame is given by  
\begin{align}
\tilde{\rho}_{2}(t)&=\text{tr}_{S_{1}}\left\{\mathcal Te^{\int_{0}^{t}dt^{\prime}\mathcal D[L(t^{\prime})]}(\rho(0))    \right\}.
\end{align}
The time order exponential can be written as 
\begin{align}
\mathcal Te^{\int_{0}^{t}dt^{\prime}\mathcal D[L(t^{\prime})]}(\cdot)=\text{id}(\cdot)+\int_{0}^{t}dt_{1}\mathcal D[L(t_{1})]+\int_{0}^{t}dt_{1}\int_{0}^{t_{1}}dt_{2}\mathcal D[L(t_{1})]\circ\mathcal D[L(t_{2})]+\cdots
\end{align}
noting that $\mathcal D[A_{1}]$ does not effect system $S_{2}$ so that with 
\begin{align} 
\mathcal D[L(t_{1})]\circ\cdots \circ \mathcal D[L(t_{n})](\rho(0))= (\mathcal D[B_{2}(t_{1})]+\tilde{\mathcal H}_{t_{1}})\circ\cdots \circ  (\mathcal D[B_{2}(t_{n})]+\tilde{\mathcal H}_{t_{n}})(\rho(0)),
\end{align}
we have 
\begin{align}
\tilde{\rho}_{2}(t)=\text{tr}_{S_{1}}\left\{\mathcal Te^{\int_{0}^{t}dt^{\prime}(\gamma^{-1}\mathcal D[B_{2}(t^{\prime})]+\tilde{\mathcal H}_{t^{\prime}})}(\rho(0))\right\}.
\end{align}
Defining $\Lambda_{t}(\cdot)=\mathcal Te^{\int_{0}^{t}dt^{\prime}\mathcal L_{t^{\prime}}}(\cdot)$ with $\mathcal L_{t}(\cdot)=\gamma^{-1}\mathcal D[B_{2}(t)](\cdot)+\tilde{\mathcal H}_{t}(\cdot)$ we then find 
\begin{align}
\Vert\rho_{2}(t)-\ket{\psi_{G}}\bra{\psi_{G}} \Vert_{1}=\Vert\ \tilde{\rho}_{2}(t)-\ket{\tilde{\psi}_{G}}\bra{\tilde{\psi}_{G}}\Vert_{1}&=\Vert\text{tr}_{S_{1}}\{(\Lambda_{t}-\tilde{\mathcal U}_{t})(\rho(0))\} \Vert_{1} \nonumber  	\\
&\leq \Vert (\Lambda_{t}-\tilde{\mathcal U}_{t})(\rho(0)) \Vert_{1} \nonumber \\
&=\Vert (\tilde{\mathcal U}_{t}^{\dagger}\circ\Lambda_{t}-\text{id})(\rho(0))\Vert _{1} 
\end{align}
where we have used that the one norm $\Vert\cdot \Vert_{1}$ is unitarily invariant and $\Vert \text{tr}_{S_{1}}\{\cdot\}\Vert_{1}\leq \Vert \cdot \Vert_{1}$. In general, the integration of 
\begin{align}
\frac{d}{dt}((\tilde{\mathcal U}_{t}^{\dagger}\circ\Lambda_{t})(\rho))=(\tilde{\mathcal U}_{t}^{\dagger}\circ \tilde{\mathcal H}^{\dagger}_{t}\circ \Lambda_{t}+\tilde{\mathcal U}_{t}^{\dagger}\circ\mathcal L_{t}\circ \Lambda_{t})(\rho)	
\end{align}
yields 
\begin{align}
\Vert (\tilde{\mathcal U}_{t}^{\dagger}\circ \Lambda_{t}-\text{id})(\rho)\Vert_{1}=\left\Vert \int_{0}^{t}dt^{\prime}\,\tilde{\mathcal U}_{t^{\prime}}^{\dagger}\circ (\tilde{\mathcal H}_{t^{\prime}}^{\dagger}+\mathcal L_{t^{\prime}})\circ \Lambda_{t^{\prime}}(\rho)    \right\Vert_{1}. 
\end{align}
such that we arrive at 
\begin{align}
	\Vert\rho_{2}(t)-\ket{\psi_{G}}\bra{\psi_{G}} \Vert_{1} &\leq \left\Vert\int_{0}^{t} dt^{\prime}\,\tilde{\mathcal U}_{t^{\prime}}^{\dagger}\circ (\tilde{\mathcal H}_{t}^{\dagger}+\mathcal L_{t^{\prime}})\circ \Lambda_{t^{\prime}}(\rho(0))  \right\Vert_{1} \nonumber \\
	&\leq \int_{0}^{t}dt^{\prime}\, \Vert \tilde{\mathcal U}_{t^{\prime}}^{\dagger}\circ (\tilde{\mathcal H}_{t^{\prime}}^{\dagger}+\mathcal L_{t^{\prime}})\circ \Lambda_{t^{\prime}}(\rho(0)) \Vert_{1} \nonumber \\
	& \leq \int_{0}^{t}dt^{\prime}\, \Vert \tilde{\mathcal U}_{t^{\prime}}^{\dagger}\circ(\tilde{\mathcal H}_{t^{\prime}}^{\dagger}+\mathcal L_{t^{\prime}})\Vert_{\infty}\nonumber\\
	&\leq \int_{0}^{t}dt^{\prime}\, \Vert \tilde{\mathcal H}_{t^{\prime}}^{\dagger}+\mathcal L_{t^{\prime}}  \Vert_{\infty}, 
\end{align}
where we used the triangle inequality, again unitary invariance and $\Vert S(\rho)\Vert_{1}\leq \Vert S\Vert_{\infty}\Vert \rho\Vert_{1} $ valid for some super operator $S$ with $\Vert\cdot \Vert_{\infty}$ being the standard operator norm. Since $\tilde{\mathcal H}_{t}^{\dagger}=-\tilde{\mathcal H}_{t}$ we find with the matrix representation of $\mathcal D[B_{2}(t)]$, obtained from row vectorization of the density operator,
\begin{align}
	\Vert\rho_{2}(t)-\ket{\psi_{G}}\bra{\psi_{G}} \Vert_{1}\leq \frac{t}{\gamma}(\Vert B\Vert_{\infty}^{2}+\Vert B^{2}\Vert_{\infty}).
\end{align}
Again, we have used unitary invariance, particularly $\Vert \tilde{B}(t)\Vert_{\infty}=\Vert B\Vert_{\infty}$, and since the fidelity error $\epsilon$ is upper bounded by $\frac{1}{2}\Vert\rho_{2}(t)-\ket{\psi_{G}}\bra{\psi_{G}}\Vert_{1}$ we have arrived at the desired result \eqref{eq:upperboundfiderror}.

\end{document}